# An Efficient Detection Mechanism for Distributed Denial of Service (DDoS) Attack

Saravanan Kumarasamy and Dr.R.Asokan

*Abstract*— Denial of Service (DoS) and Distributed Denial of Service (DDoS) attacks have emerged as a popular means of causing collection particular overhaul disruptions, often for total periods of instance. The relative ease and low costs of initiation such attacks, supplemented by the present insufficient sate of any feasible defense method, have made them one of the top threats to the Internet centre of population nowadays. Since the rising attractiveness of web-based applications has led to quite a lot of significant services being provided more than the Internet, it is very important to monitor the network transfer so as to stop hateful attackers from depleting the assets of the network and denying services to rightful users. The most important drawbacks of the presently existing defense mechanisms and propose a new-fangled mechanism for defending a web-server against a DDoS attack. In the proposed mechanism, incoming traffic to the server is always monitored and some irregular rise in the inbound traffic is without delay detected. The detection algorithm is based on a statistical analysis of the inbound traffic on the server and a robust suggestion testing structure. While the detection procedure is on, the sessions from the rightful sources are not disrupted and the load on the server is restored to the usual level by overcrowding the traffic from the attacking sources. The accurate modules employ multifaceted detection logic and hence involve additional overhead for their execution. On the other hand, they have very huge detection accuracy. Simulations approved on the proposed mechanism have produced results that show efficiency of the proposed defense mechanism against DDoS attacks.

## I. INTRODUCTION

A *denial of service* (DoS) attack is defined as an unambiguous effort by a horrible user to get through the capital of a server or a network, thereby preventing rightful users from availing the services provided by the system. The most ordinary DoS attacks typically engage flooding with a huge volume of travel and overwhelming network resources such as bandwidth, buffer space at the routers, CPU time and revival cycles of the end server. A small number of the common DoS attacks are SYN flooding, UDP flooding, DNS based flooding, ICMP directed broadcast, Ping flood attack, IP fragmentation, and CGI attacks [1]. Based on the number of violent equipment deployed to perform the attack, DoS attacks are classified into two broad categories: (i) The single intruder consume with the maximum possible bandwidth obtain from the packets from the one individual users machine or (ii) The various frequent attackers are coordinator with the same effect from multiple machine with a different network. The DDoS attacks identification process is very tricky to notice. It is extremely important that suitable defense mechanism should be in place to notice such attacks as speedily as possible.

In this paper, a robust method is proposed to care for a web server from DDoS attack utilizing some easily reachable information in the server. Through the whole system is not promise to detect the DDoS attack fully and also server within short duration time server quickly shutdown appears, on the same time normal network service cannot possible. The new efficient detection algorithm is used to find the flexible solution for the DDoS Attacks. The efficient detection algorithm is classified different modules. Find out the DDoS attack is very quickly with simple statistical analysis of the network traffic and also very less computational memory overhead on the server. The development of the algorithm is based on the legitimate users are allowed as well as attackers are identified and also deleted. This aspect of behaviour DDoS attacks is not taken into account in numerous of the commercial solutions [3].

## II. DISTRIBUTED DENIAL OF SERVICE (DDoS) ATTACK

There are two most significant types of DDoS attacks [4]. The attacks of the first type's attempt to get through the resources of the injured party host. In general the victim is a web server or proxy connected to the Internet. When the traffic load becomes enormously eminent, the victim host starts dropping packets both from the authentic users and attack sources. The victim also sends message to all the sources to trim down their sending rates. The authentic sources slowly downwards their rates while the attack sources still maintain or increase their sending rates. Eventually, the victim host's resources, such as CPU cycles and memory space get exhausted and the victim is unable to service its genuine clients. The attacks of the second type target network bandwidth. If the malicious traffics in the network are able to dominate the communication links, then traffics from the genuine sources are affected. The effects of bandwidth DDoS attacks are usually more severe than the resource consumption attacks. In this section, some classic bandwidth attacks are discussed.

The *SYN flood* attack occur means server needs to provide the huge memory data structure for authentication of incoming SYN packets. During SYN flood attacks, the attacker sends more number of SYN packets with source addresses. In the request response process time, when the server sends the request information into the memory stack, it will wait for the verification from the client that sends the request. Thus the request is waiting to be established, it will stay in the memory stack. Since the source addresses used in SYN flood attacks may be spurious, the server will not receive confirmation packets for requests created by the SYN flood attack. Each half-open connection will stay on the memory stack in anticipation of it times out. This causes the memory stack getting occupied. Furthermore, including genuine requests can be processed and the services of the system are disabled. SYN floods remain one of the most powerful flooding methods.

The *smurf* attack is a type of ICMP flood, where attackers use ICMP echo request packets directed to IP broadcast addresses from remote locations to generate denial of service attacks. There are three entities in these attacks: the attacker, the intermediary, and the victim. First, the attacker sends one ICMP echo request packet to the network broadcast address and the request is forwarded to all the hosts within the intermediary network. Second, all of the hosts within the intermediary network send the ICMP echo replies to flood the victim. Solutions to the smurf attack include disabling the IP-directed broadcast service at the intermediary network. Nowadays, smurf attacks are quite rare in the Internet since defending against such attacks are not difficult.

An *HHTP flood* refers to an attack that bombards web servers with HTTP requests. HTTP flood is a common feature in most botnet software. To send an HTTP request, a valid TCP connection has to be established, which requires a genuine IP address. Attackers can achieve this by using a bot's IP address. Moreover, attackers can craft the HTTP requests in different ways in order to either maximize the attack power or avoid detection. For example, an attacker can instruct the botnet to send HTTP requests to download a large file from the target. The target then has to read the file from the hard disk, store it in memory, load it into packets and then send the packets back to the botnet. Hence, a simple HTTP request can incur significant resource consumption in the CPU, memory, input/output devices, and outbound Internet link.

Another important DDoS attack is the *SIP flood* attack. A widely supported open standard for call setup in the *voice over IP* (VoIP) is the *session initiation protocol* (SIP) [5]. Generally, SIP proxy servers require public Internet access in order to accept call setup requests from any VoIP client. Moreover, to achieve scalability, SIP is typically implemented on top of UDP in order to be stateless. In one attack scenario, the attacker can flood the SIP proxy with many SIP INVITE packets that have spoofed source IP addresses [6]. To avoid any anti-spoofing mechanisms, the attackers can also launch the flood from a botnet using non-spoofed source IP addresses. There are two categories of victims in this attack scenario. The first types of victims are the SIP proxy servers. Not only will their server resources be depleted by processing the SIP INVITE packets, but their network capacity will also be consumed by the SIP INVITE flood. In either case, the SIP proxy server will be unable to provide VoIP service. The second types of victims are the call receivers. They will be overwhelmed by the forged VoIP calls, and will become nearly impossible to reach by the genuine callers.

## III. RELATED WORK

Protection against DoS and DDoS attacks extremely depends on the model of the network and the type of attack. Several mechanisms have been proposed to solve the problem. However, most of them have weakness and fail under positive scenarios. In this section, some of the existing defense mechanisms against DoS and DDoS attacks are discussed briefly.

*Protocol reordering* and *Protocol enhancement* methods make security protocols more robust and less susceptible to resource consumption attacks [11][12].

*Network ingress filtering* is a mechanism proposed to prevent attacks that use spoofed basis addresses [13]. This involves configuring the routers to drop packets that have illegitimate source IP addresses. One of the serious pitfalls of this way is its inability to curtail a flood attack that originates with a spoofed IP address from within the network.

*ICMP traceback* messages are useful to recognize the path taken by packets through the Internet [14]. This requires a router to use a very low chance with which traceback messages are sent along with the traffic. Hence, with adequately large number of messages, it is possible to conclude the route taken by the traffic during an attack. This enables localization of the aggressive host.

An approach to conquer the problems connected with ascertaining the validity of IP addresses in ingress filtering is to use the routing in sequence instead of just the source address. *IP traceback* proposes a reliable way to perform hop by hop tracing of a packet to the attacking source from where it originated [15][16]. However, this requires coordinated effort from all the routers in the network along with the path from the victim to the attacker, and examination of the packet logs.

*Deterministic packet marking* (DPM) is another device to detect DoS attacks [17]. It relies on routing information inscribed in the packet header by the routers as the packet traverses the network. This come near leads to an increase in the size of the IP packet header as the size of IP header increases linearly with the number of hops traversed. The consequential variable header size increases the complexity of processing.

*Probabilistic packet marking* (PPM) for IP traceback is a method that attempts to get better DPM [18]. It eliminates IP address spoofing by allowing each router to probabilistically inscribe local path information onto a packet that traverses it [17]. This enables a victim host to localize the attacking source while retaining fixed sized packet headers. The mechanism is dependent on route stability between the attacker and the victim to localize the attacker. A similar mechanism known as *route-based packet filtering* has been proposed in [19], which uses the source and destination addresses on a packet header to ascertain the strength of the route.

Yaar et al. have proposed an approach, called *path identifier* (Pi), in which a trail fingerprint is embedded in each packet, enabling a victim to identify packets traversing the same paths through the Internet on a per packet basis, regardless of source IP address spoofing [20]. Pi allows the victim to take a practical role in defending against a DDoS attack by using the Pi mark to filter out attack packets.

*Pushback* approaches have been proposed to extract attack signatures by rate-limiting the doubtful traffic destined to a congested link [21][22]. Since the DDoS flooding travel does not follow the end-to-end flow control protocol in the path, it is possible to find the congestion signature using the packet drop statistics. Pushback differentiates attacking travel from rightful travel by monitoring whether the suspicious travel obeys the end-to-end congestion control.

Gil et al. have proposed a scheme named *MULTOPS* [23] in which routers notice bandwidth attacks using a heuristic based on packet sending rates. Under non-attack circumstances, the packet flow rate in the way over the Internet is directly proportional to the packet flow rate in the differing direction. As soon as this condition is violated, an attack is supposed to have occurred. However, efficiency of MULTOPS degrades with randomized IP source addresses.

Mirkovic et al. have proposed a scheme named D-WARD that performs statistical traffic profiling at the edge of the networks to notice new types of DDoS attacks [24]. By monitoring the nominal per-destination type traffic arrival and departure rates of TCP, UDP, ICMP packets, and on observing any irregular asymmetric behavior of the two-way traffic at the edge router connecting to a stub-network, D-WARD aims at stopping DDoS attacks near their sources.

Zou et al. have presented an adaptive defense scheme that adjusts its configurations according to the network conditions and attack severity in order to minimize the joint cost introduced by false positives (wrongly identify normal attack as an attack) and fake negatives (wrongly identify attack traffic as normal) [25].

*Client side puzzle* and other *pricing algorithms* [26][27][28] are efficient tools to make protocols less vulnerable to depletion attacks of processing power. However, in case of disseminated attacks their effectiveness is debatable.

## IV. ALGORITHMS IN THE INTERFACE MODULE

In this DDOS detection and protection system two types of algorithm,
- Special flow Monitoring Algorithm
- IP Traceback Algorithm

The Special flow Monitoring Algorithm

**The attack packets are reaching a current router as follow a1; a2; . . .; ak.**
**Attack packet rates are k.**
**Set count as each flow $x_1, x_2, \ldots, x_n$.**
And the p represent **distribution rate of packets.**

$$p_i = x_i \sum_{i=1}^{n} x$$

**Identify the flows $f_1, f_2, \ldots, f_n$.**
**The entropy variation formula is,**

$$H(F) = -\sum_{i=1}^{n} (p_i \, log \, p_i)$$

The special flow monitoring algorithm is running at the non-attack period, accumulating information from normal network flows, and progressing the mean and the standard variation of flows. The progressing suspends when a DDoS attack is ongoing. Once a DDoS attack has been confirmed by any of the existing DDoS detection algorithms, then the victim starts the IP traceback algorithm. This continuously monitoring the http request from the internet. When the request is coming, it identifies the IP address and stored in cache and starts counting the request from the same IP address and also maintains the timer.

Initiate the local parameter X,U,D.
U={$u_i$} be set of upstream routers, D={$d_i$} be set
a destination address of the packet and the victim is V.
The attack flow as, $f_i = <u_i, V>$, i=1,2,..n. That's like a data flow as $f_1, f_2, \ldots, f_n$.
For i=1 to n
{
Calculate H(F\$f_i$)
If H(F\$f_i$)>$x_n$
Upstream router of $f_i$ to set A
else break;

```
end if;
end for;
}
```

More than 20 requests within one second from same IP address is considered as DDOS attack. Then the IP address is blocked for certain time periods prevention that means the suspicious IP address is blocked for certain time periods. That's like a monitoring process are very effective to monitoring the network and this monitoring is used to find out the attacker easily. The monitoring process is used to pushback when the attack is occurring. The traceback process is find out the attacker from the network when the attack traffic is present in the network

In the IP traceback algorithm is installed at routers. It is initiated by the victim, and at the upstream routers, it is triggered by the IP traceback requests from the victim or the downstream routers which are on the attack path. The proposed algorithms are independent from the current routing software, they can work as independent modules at routers. As a result, we do not need to change the current routing software.

## V. SIMULATIONS AND RESULTS

The simulation is implemented in Network Simulator 2.31[19], a simulator for wired networks. The simulation parameters are provided in Table 1. We implement the random waypoint movement model for the simulation, in which a node starts at a random position, waits for the pause time, and then moves to another random position with a velocity chosen 35 m/s. A packet size of 512 bytes and a transmission rate of 4 packets/s,

**Performance Metrics:** In our simulations utilize several performance metrics to compare the proposed web server model with the existing one [20]. The following metrics were measured for the comparison were

*a)* Throughput: Number of packets sends in per unit of time.

*b)* Packet delivery fraction *(PDF):* The ratio between the numbers of packets sends by source nodes to the number of packets correctly received by the corresponding destination nodes.

*c)* End to End delay: - Measure as the average end to end latency of data packets.

*d)* Normalized routing load: Measured as the number of routing packets transmitted for each data packet delivered at the destination.

**TABLE 1 Simulation Parameters for Case Study**

| | |
|---|---|
| Number of nodes | 13 |
| Dimension of simulated area | 800×600 |
| Simulation time (sec) | 35 |
| Radio range | 250m |
| Traffic type | CBR, 3pkts/s |
| Packet size (bytes) | 512 |
| Number of traffic connections | TCP/UDP |
| Maximum Speed (m/s) | 35 |
| Node movement | random |
| Types of attack | DDOS |

Poisson model of travel arrival is chosen as it is particularly suitable for dealing with some Internet protocols if its parameters are set appropriately. *Internet control message protocol* (ICMP), *network time protocol* (NTP) and *domain name service* (DNS) clients post many small packets of constant size with uniformly distributed inter-packet arrival time. These protocols resemble very intimately to the assumptions that have been made in the simulation. This makes the results in simulation realistic. Since in practical scenarios, the number of genuine clients that connect to a server may also vary over a broad range, the following cases are considered:

*Case 1*: For a little corporate server, the number of legal clients is low, say $N(t) = 5$. presumptuous that the capacity of the server is high, the average load on the server will be less. consequently, the number of attacking hosts should be high, say $A(t) = 40$. Hence, in this scenario, for an effective attack we must have $N(t) << A(t)$.

*Case 2*: For a server of medium size, it may be supposed that $N(t) = 50$ and a successful attacker can launch his/her attack from a fewer number of hosts. Thus it may be assumed that $A(t) = 50$ in this case. As the number of officially permitted clients and the number of attacking sources are of comparable size, it is easier for the attacker to hide his/her attack in this case. Therefore, in this situation, $N(t) \_ A(t)$.

*Case 3*: For a worldwide portal server, there can be a very large number of legal clients, say $N(t) = 10000$. In this circumstances, it is not possible for that attacker to easily estimate the required number of attacking hosts. In this case, it is assumed that the attacker chooses a reasonably high value of $A(t)$, say $A(t) = 5000$, and opts for a very high attacking rate: $\_a = \_n*10$. Therefore, in this case: $N(t) > A(t)$.

In the first simulation, a huge number of hosts are taken to test the efficiency of the proposed mechanism on a large system. The simulation parameters are listed in Table 2.

**Table 2. Simulation parameters for Simulation I**

| Parameter | Value |
|---|---|
| Number of legal clients ($N(t)$) | 12000 |
| Number of attacking hosts ($A(t)$) | 6000 |
| Mean normal traffic rate ($\_n$) | 0.4 |
| Mean attack traffic rate ($\_a$) | 0.5 |

**TABLE 3 Overall summaries of Results in all Cases**

| Parameter | Normal Case | Attack Case |
|---|---|---|
| SEND | 928 | 633 |
| RECEIVE | 924 | 582 |
| ROUTING PACKETS | 169 | 229882 |
| PACKET DELIVRY FRACTION | 95.1 | 89.43 |
| THROUGHPUT | 107 | 58 |
| NORMAL ROUTING LOAD | 0.12 | 456.19 |
| AVERAGE END TO END DELAY | 852.04 | 751.64 |
| No. Of dropped data(packets) | 23 | 51 |
| No. Of dropped data(bytes) | 23852 | 44556 |

According to act analysis in normal case, in attack case we monitor that DDOS attack definitely pretentious the network and our scheme is successfully defence the network and also provides the safety against them. In case of attack we scrutinize that the routing load is very high because attacker node are constantly transmit the routing packets to their neighboured and every node in network are reply to attacker node by that heavy congestion is occur. Packet delivery fraction and end to end delay are also goes near to the ground, which shows that packets are not deliver exactly and number of dropped data is goes high approximately twice to the usual condition.

## VI. CONCLUSION

The steady progress of DDoS attacks as a means for achieving political, economic and commercial gains, and the relative ease, low costs, and limited responsibility in launching such attacks, have rendered them one of the top pressure to today's Internet services. Although a diversity of independent DDoS attack prevention, mitigation, and traceback techniques have been proposed by researchers over the last decades, their relative uptake has been minimal at beast, due to the be short of a robust, foolproof, and universal DDoS attack defense mechanism. In this paper, a mechanism is presented for detection and anticipation of DDoS attacks on a web server while the proposed mechanism does not affect the traffic from genuine clients, it effectively blocks traffic from the attack sources with a very low false positive rate and high detection accuracy. The simulation results demonstrate the efficiency of the proposed mechanism. Development of an analytical framework for finding an optimum value of the traffic analysis window ($ws$) and design of a heuristic for faster attack discovery with more accuracy are the two areas in which future research work will be approved out.

**Saravanan Kumarasamy**

received the M.E degree 2008 in computer science from Dr.MCET, Anna University, Chennai, India. He is currently working as a Lecturer at the Faculty of Engineering, Erode Sengunthar Engineering College, Erode, Tamilnadu. He has published 5 paper in International Journal, 10 papers in National Conference and 02 papers in International Conference..His current research interests are information security, computer communications and DDoS Attacks. He is currently pursuing Ph.D. under Anna University of Technology, Coimbatore.

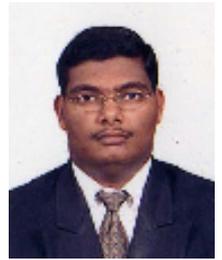

**Dr.R.Asokan**

received the Ph.D degree in Information and Communication Engineering from Anna University, Chennai. He has 25 years of teaching experience. At present he is working as Principal at Kongunadu College of Engineering and Technology, Thottiam, Trichy. He has published more than 70 papers in National and International Journals and Conferences. He has organized more than 20 Seminars, Workshops and Conferences.

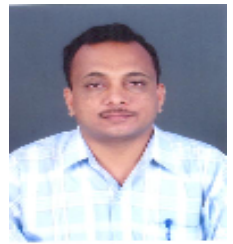

He has delivered around 40 special lectures in various summer / winter school/ sponsored programmes. He is the associate editor for Journal of selected areas in telecommunication and also Editorial board member for five International Journals.  His areas of interest include communication networks, network security and image processing.  He is an active member in many Professional bodies like IETE, ISTE, CSI, ACS etc.